\documentclass[12pt]{article}

\usepackage{amsmath,amssymb,amsfonts,amsbsy}
\usepackage{cite}
\usepackage{graphicx}
\usepackage{wrapfig}
\usepackage[font=small,labelfont=bf,labelsep=period]{caption}

\begin{document}
\addcontentsline{toc}{subsection}{{$O(\alpha_s)$ spin effects in
  $e^+e^-\to q\bar q (g)$}\\
{\it J.G.~K\"orner}}

\setcounter{section}{0}
\setcounter{subsection}{0}
\setcounter{equation}{0}
\setcounter{figure}{0}
\setcounter{footnote}{0}
\setcounter{table}{0}

\begin{center}
\textbf{$O(\alpha_s)$ SPIN EFFECTS IN $e^+e^-\to q\bar q \,(g)$}

\vspace{5mm}

S.~Groote$^{\,1,\,2}$ and \underline{J.G.~K\"orner}$^{\,1\,\dag}$

\vspace{5mm}

\begin{small}
  (1) \emph{Institut f\"ur Physik, Johannes-Gutenberg-Universit\"at,
  Mainz, Germany} \\
  (2) \emph{F\"u\"usika Instituut, Tartu \"Ulikool, Tartu, Estonia} \\
  $\dag$ \emph{E-mail: koerner@thep.physik.uni-mainz.de}
\end{small}
\end{center}

\vspace{0.0mm} 

\begin{abstract}
We discuss $O(\alpha_s)$ spin effects in $e^+e^-\to q \bar{q}\,(g)$ for
the polarization of single quarks and for spin-spin correlations of the final 
state quarks.
Particular attention is paid to residual mass effects in the limit
$m_{q} \to 0$ which are described in terms of universal helicity flip
and helicity non-flip contributions.
\end{abstract}

\vspace{7.2mm} 

\section{Introduction}
In a series of papers we have investigated $O(\alpha_s)$ final state spin
phenomena in $e^+e^-\to q\bar q\,(g)$. In 
\cite{Korner:1993dy,Groote:1995yc,Groote:1996nc} we have provided analytical 
results for the polarization of single quarks and in 
\cite{Groote:1997su,Groote:2009zk}
for longitudinal spin-spin correlations of the final state quarks including 
their dependence on
beam-event correlations \footnote{Numerical $O(\alpha_s)$ results on final
state quark polarization effects in $e^+e^-\to q \bar{q}\,(g)$ can be found
in \cite{Brandenburg:1998xw}.}. By carefully taking the $m_q\to 0$ limits of our
analytical $O(\alpha_s)$ results we have found that, at $O(\alpha_s)$, the 
single-spin polarization $P^{(\ell_1)}$ and the spin-spin polarization 
correlation $P^{(\ell_1\ell_2)}$
do not agree with their $m_q=0$ counterparts. In fact, averaging over
beam-event correlation effects, at $O(\alpha_s)$ one has 
\begin{equation}\label{anpol}
P^{(\ell_1)}_{m_q\to 0}=P^{(\ell_1)}_{m_q=0}
\left(1-\Big[\frac23\,\frac{\alpha_s}\pi\Big]\right),\qquad
P^{(\ell_1\ell_2)}_{m_q\to 0}=P^{(\ell_1\ell_2)}_{m_q=0}
\left(1-\Big[\frac43\,\frac{\alpha_s}\pi\Big]\right)
\end{equation}
where the residual $m_q\to 0$ contributions are encased in square brackets.
In Eq.(\ref{anpol}) $P^{(\ell_1)}$ denotes the longitudinal single-spin 
polarization of the final state quark and $P^{(\ell_{1}\ell_{2})}$ denotes 
the longitudinal 
spin-spin polarization correlation of the quark and antiquark. From 
Eq.(\ref{anpol}) it
is apparent that at $O(\alpha_{s})$\, $QCD(m_{q}=0)\neq QCD(m_{q}\to 0)$ in
polarization phenomena. We mention that the longitudinal 
polarization components are the only polarization components that survive in 
the high energy (or $m_{q}=0$) limit. Here and in the following 
the residual mass contributions will be called anomalous
contributions for the reason that the anomalous helicity
flip contribution enters as absorptive input in the dispersive 
derivation of the value of the axial anomaly \cite{Dolgov:1971ri}. We shall 
see further on 
that $ P^{(\ell_{1})}_{m_{q}= 0}=g_{14}/g_{11}$ where $g_{14}$ and
$g_{11}$ are electroweak coupling coefficients and 
$P^{(\ell_{1}\ell_{2})}_{m_{q}= 0}=-1$ independent of the electroweak coupling
coefficients. It is important to keep in mind that there are no residual mass
effects at $O(\alpha_{s})$ in the unpolarized rate.

By an explicit calculation we have checked that the 
difference of the two results originates from the near-forward region
(see \cite{gk09}) which 
is very suggestive of an explanation in terms of universal near-forward 
contributions. That
there is a universal helicity flip contribution in the splitting process
$q_{\pm} \to q_{\mp} + g$ has been noted some time ago in the
context of QED \cite{Lee:1964is} (see also \cite{Smilga:1990uq,Falk:1993tf}). 
We have found that
there also exists a universal helicity non-flip contribution 
$q_{\pm} \to q_{\pm} + g$ which is perhaps not so well known in the 
literature. In fact,
the anomalous contribution to the single-spin polarization $P^{(\ell_{1})}$ 
results to
$100 \%$ from the universal helicity non-flip contribution whereas the 
anomalous contributions to the spin-spin correlation $P^{(\ell_{1}\ell_{2})}$ 
is $50\%$ helicity flip and $50\%$ helicity non-flip.

\section{Definition of polarization observables}
In the limit $m_{q}\to0$ the relevant spin degrees of freedom are the
longitudinal polarization components $s_{1}^{\ell}=2\lambda_{q}$ and 
$s_{2}^{\ell}=2\lambda_{\bar{q}}$ of the quark and the antiquark. 
One defines an unpolarized structure function $H$ and single--spin
and spin--spin polarized structure functions $H^{(\ell_1)}$, $H^{(\ell_2)}$ 
and $H^{(\ell_1\ell_2)}$, resp.,  according to
\begin{equation}\label{spinspin1}
H(s_{1}^{\ell}s_{2}^{\ell})=\frac14\left(H
  +H^{(\ell_1)}s_{1}^{\ell}
  +H^{(\ell_2)}s_{2}^{\ell}
  +H^{(\ell_1 \ell_{2})}s_{1}^{\ell}s_{2}^{\ell}\right)\,.
\end{equation} 
Eq.(\ref{spinspin1}) can be inverted to give
\begin{eqnarray}\label{uparrow1}
H&=&\big[H(\uparrow \uparrow)\big]
  +H(\uparrow \downarrow)+H(\downarrow \uparrow)
  +\big[H(\downarrow \downarrow)\big],\\
H^{(\ell_1)}&=&\big[H(\uparrow \uparrow)\big]
  +H(\uparrow \downarrow)-H(\downarrow \uparrow)
  -\big[H(\downarrow \downarrow)\big],\nonumber\\
H^{(\ell_2)}&=&\big[H(\uparrow \uparrow)\big]
  -H(\uparrow \downarrow)+H(\downarrow \uparrow)
  -\big[H(\downarrow \downarrow)\big],\nonumber\\
H^{(\ell_1\ell_2)}&=&\big[H(\uparrow \uparrow)\big]
  -H(\uparrow \downarrow)-H(\downarrow \uparrow)
  +\big[H(\downarrow \downarrow)\big].\nonumber
\end{eqnarray}
We have indicated in (\ref{uparrow1}) that the spin configurations
$H(\uparrow \uparrow)$ and $H(\downarrow \downarrow)$ can only be populated
by anomalous contributions in the $m_{q} \to 0$ limit.
The normalized single-spin polarization $P^{(\ell_{1})}$ and the spin--spin 
correlation 
$P^{(\ell_{1}\ell_{2})}$ are then given by 
\begin{equation}
P^{(\ell_{1})}=\frac{g_{14}}{g_{11}}\,\frac{H^{(\ell_{1})}}{H}
\qquad P^{(\ell_{1}\ell_{2})}=\frac{H^{(\ell_{1}\ell_{2})}}{H}\,,
\end{equation}
where $g_{14}$ and $g_{11}$ are $q^{2}$--dependent electroweak coupling 
coefficients (see e.g. \cite{Groote:1996nc}). For the ratio of electroweak
coupling coefficients one finds $g_{14}/g_{11}= -0.67$ and -0.94 for 
up-type and down-type quarks, respectively, on the $Z_{0}$ resonance,
and $g_{14}/g_{11}= -0.086$ and -0.248 for $q^{2} \to \infty$.

Considering the fact that one has 
$H(\uparrow \uparrow)=H(\downarrow \downarrow)=0$ for
$m_{q}=0$, $H^{pc}(\uparrow \downarrow)=H^{pc}(\downarrow\uparrow~\!\!\!)$ for 
the p.c. structure functions $(H,H^{(\ell_1\ell_2)})$, and
$H^{pv}(\uparrow \downarrow)=-H^{pv}(\downarrow \uparrow)$ for the p.v. 
structure function $H^{(\ell_1)}$ with $H^{pv}(\uparrow \downarrow)
=H^{pc}(\uparrow \downarrow)$, it is not difficult to see from 
Eq.(\ref{uparrow1}) that 
$H^{(\ell_{1})}/H=+1$ and $H^{(\ell_{1}\ell_{2})}/H=-1$ for $m_{q=0}$
to all orders in perturbation theory.

\section{Polarization results for $m_{q}\to 0$ and $m_{q}=0$}
The results of taking the $m_{q}\to 0$ limit of our analytical finite mass
results in
\cite{Korner:1993dy,Groote:1995yc,Groote:1996nc,Groote:1997su,Groote:2009zk}
can be concisely written as 
\begin{eqnarray}\label{hadfun}
H^{pc}(s_1^\ell,s_2^\ell)
  \!\!\!&=&\!\!\!\frac14\Big(H^{pc}+H^{pc\,(\ell_1\ell_2)}s_1^\ell s_2^\ell\Big)
  =N_cq^2\left((1-s_1^\ell s_2^\ell)\left(1+\frac{\alpha_s}{\pi}\right)
  +\left[\frac43\times\frac{\alpha_s}\pi s_1^\ell s_2^\ell\right]\right),
  \nonumber\\
H^{pv}(s_1^\ell,s_2^\ell)
  \!\!\!&=&\!\!\!\frac14\Big(H^{pv(\ell_1)}s_1^\ell+H^{pv(\ell_2)}s_2^\ell\Big)
  =N_cq^2(s_1^\ell-s_2^\ell)\left(1+\frac{\alpha_s}\pi
  -\left[\frac23\times\frac{\alpha_s}\pi\right]\right).
\end{eqnarray}
We have again indicated the parity nature of the structure functions
((pc): parity conserving; (pv): parity violating).
The $m_{q}=0$ results are obtained from
Eq.(\ref{hadfun}) by dropping the anomalous square bracket 
contributions (see \cite{Groote:1996nc,Groote:2009zk}). 

It is not difficult to see that one has $H^{pc}=-H^{pc\,(\ell_1\ell_2)}
=H^{pv(\ell_1)}=-H^{pv(\ell_2)}$ for $m_{q}=0$ by commuting $\gamma_{5}$ 
through
the relevant $m_{q}=0$ diagrams. As Eq.(\ref{hadfun}) shows these relations 
no longer hold true for the anomalous contributions showing again that the
anomalous contributions originate 
from residual mass effects which obstruct the simple 
$\gamma_{5}$-commutation structure of the $m_{q}=0$ contributions. 

\section{Near-forward gluon emission }
In Table 1 we list the helicity amplitudes 
$h_{\lambda_{q_{1}}\lambda_{q_{2}}\lambda_{g}}$ for the splitting process
$q(p_{1}) \to q(p_{2}) +g(p_{3})$ and the $\cos\theta$--dependence of the helicity amplitudes in 
the near-forward region. We also list the 
difference of the initial helicity and the final helicities
$\Delta\lambda=\lambda_{q_{1}}-\lambda_{q_{2}}-\lambda_{g}$.
\begin{center}
\begin{tabular}{lcc} \hline\hline 
\\[-0.3cm]
${\textstyle h_{\lambda_{q_{1}}\lambda_{q_{2}}\lambda_{g}}}\qquad$
&${\Delta\lambda}$
&\qquad$\cos\theta$ {\mbox dependence}\\
\\[-0.3cm]
 \hline
\\[-0.3cm]
$h_{1/2\,\,1/2\,\,+1}\qquad$&--1&$\sim \sqrt{(1-\cos\theta)}$\\
\vspace{0.1cm}
$h_{1/2\,\,1/2\,\,-1}\qquad$&+1&$\sim \sqrt{(1-\cos\theta)}$\\
$h_{1/2\,\,-1/2\,\,+1}\qquad$&0&$\sim m_{q}/E$\\
\\[-0.4cm]
$h_{1/2\,\,-1/2\,\,-1}\qquad$&+2&$0$\\
\\[-0.3cm]\hline\hline
\end{tabular}
\end{center}
\begin{center}
Table 1. Helicity amplitudes and their near-forward
behaviour.\\
\hspace{-1.0cm}Column 2 shows helicity balance
$\Delta\lambda=\lambda_{q_{1}}-\lambda_{q_{2}}-\lambda_{g}$.
\end{center}
In the forward direction $\cos\theta=1$ the only surviving helicity
amplitude is $h_{1/2\,\,-1/2\,\,+1}$ for which the helicities satisfy the 
collinear angular momentum conservation rule $\Delta \lambda=0$.  

Squaring the helicity flip amplitude $h_{1/2\,\,-1/2\,\,+1}$ and adding in 
the propagator denominator
factor $2p_{2}p_{3}=2E^{2}x(1-x)\big(1-\sqrt{1-m^{2}/E^{2}}\,\big)$ one 
obtains 
$(x=E_{3}/E$; $E$ is the energy of the incoming quark and $s=4E^{2}$)
\begin{equation}
\label{flip}
\frac{d\sigma_{[hf]}}{dx \,dcos\theta}= 
\sigma_{Born}(s)\,C_{F}\frac{\alpha_{s}}{4\pi}\,x  
\frac{m_{q}^{2}}{E^{2}}
\frac{1}{\Big(1-\cos\theta\sqrt{1-m_{q}^{2}/E^{2}}\,\,\Big)^{2}}
\end{equation}
Note that the helicity flip contribution $h_{1/2\,\,-1/2\,\,+1}$
is not seen in a $m_{q}=0$ calculation.
The helicity flip splitting function $d\sigma_{[hf]}/d\cos\theta$ is strongly 
peaked in
the forward direction. Using the small $(m_{q}^{2}/E^{2})$--approximation one
finds that $\sigma_{[hf]}$ has fallen to $50\%$ of its forward peak value at
$\cos\theta=1-(\sqrt{2}-1)m_{q}^{2}/(2E^{2})$.

Integrating Eq.(\ref{flip}) over $\cos\theta$, we obtain
\begin{equation}
\label{iflip}
\frac{d\sigma_{[hf]}}{dx}=
\sigma_{Born}(s)C_{F}\frac{\alpha_{s}}{2\pi}x\equiv \sigma_{Born}(s) 
D_{[hf]}(x)
\end{equation}
where $D_{[hf]}=C_{F}(\alpha_{s}/2\pi)x$ is called the helicity flip 
splitting function \cite{Falk:1993tf}. The integrated helicity flip 
contribution survives
in the $m_{q}\to0$ limit since the integral of the last factor in 
Eq.(\ref{flip}) is proportional to $1/m_{q}^{2}$ which cancels the
overall $m_{q}^{2}$ factor in Eq.(\ref{flip}). It is for this reason that
the survival of the helicity flip contribution is sometimes called an
$m/m$-effect. 
 
For the helicity non-flip contribution one finds
\begin{equation}
\label{nonflip}
\frac{d\sigma_{nf}}{dx \,dcos\theta} = \sigma_{Born}(s)\,
C_{F}\frac{\alpha_{s}}{2\pi}\frac{1+(1-x)^{2}}{x}
\frac{(1-\cos\theta)}
{\Big(1-\cos\theta\sqrt{1-m_{q}^{2}/E^{2}}\,\,\Big)^{2}}
\end{equation}
The helicity non-flip splitting function $d\sigma_{nf}/d\cos\theta$ 
vanishes in the forward direction but is strongly 
peaked in the 
near-forward direction.
Using again the small $(m_{q}^{2}/E^{2})$ approximation $\sigma_{nf}$ can be 
seen to peak 
at $\cos\theta=1-m_{q}^{2}/(2E^{2})$.

Integrating Eq.(\ref{nonflip}) one obtains
\begin{equation}
\frac{d\sigma_{nf}}{dx}=\sigma_{Born}(s)C_{F}
\frac{\alpha_{s}}{\pi}\frac{1+(1-x)^{2}}{x}
\ln\frac{E}{m_{q}}\equiv \sigma_{Born}(s) D_{nf}(x)
\end{equation}
where we have retained only the leading-log contribution. $D_{nf}(x)$ can be 
seen to be the usual non-flip splitting function.

We now turn to the anomalous helicity non-flip contribution. Let us rewrite
the non-flip contribution in the form
\begin{equation}
\sigma_{nf}
=\underbrace{\sigma_{nf}+\sigma_{[hf]}}_{{\displaystyle \sigma_{total}}}
-\sigma_{[hf]}
\end{equation}
By explicit calculation we have seen that the total unpolarized rate 
$\sigma_{total}$ has no anomalous
contribution. The conclusion is that there is an anomalous contribution
also to the non-flip transition with the strength 
$-D_{[hf]}=-C_{F}(\alpha_{s}/2\pi)x$. 
We conjecture that the same pattern holds true for other processes, i.e. that
there are no anomalous contributions to unpolarized rates but that there are
anomalous contributions to both helicity flip and non-flip contributions
in polarized rates. 

Taking into account the anomalous helicity flip $\sigma_{[hf]}$ and
non-flip $\sigma_{[nf]}$ contributions  
the pattern of the anomalous helicity contributions to the 
various spin configurations can then be obtained in terms of the
Born term contributions and the universal flip and non-flip contributions
$\pm \int_{0}^{1}D_{[hf]}(x)dx=\pm C_{F}\alpha_{s}/(4\pi)
= \pm \, \alpha_{s}/(3\pi)$. Using a rather suggestive notation for the
anomalous contributions one has
\begin{eqnarray}
\label{anom}
\left[\uparrow\uparrow\right]&=(\uparrow\downarrow_{[hf]})
  +(\downarrow_{[hf]}\uparrow)
  &=\Big((\uparrow \downarrow)_{\it Born}
  +(\downarrow \uparrow)_{\it Born}\Big)
\left[C_F\frac{\alpha_s}{4\pi}\right]\nonumber\\
\left[\uparrow\downarrow\right]&=(\uparrow\downarrow_{[nf]})
  +(\uparrow_{[nf]}\downarrow)  
  &=-2(\uparrow\downarrow)_{\it Born}\left[C_F\frac{\alpha_s}{4\pi}\right],
\nonumber\\
\left[\downarrow\uparrow\right]&=(\downarrow\uparrow_{[nf]})
  +(\downarrow_{[nf]}\uparrow)  
  &=-2(\downarrow\uparrow)_{\it Born}\left[C_F\frac{\alpha_s}{4\pi}\right],
\nonumber\\
\left[\downarrow\downarrow\right]&=(\downarrow\uparrow_{[hf]})
  +(\uparrow_{[hf]}\downarrow)
  &=\Big((\downarrow \uparrow)_{\it Born}
  +(\uparrow \downarrow)_{\it Born}\Big)
\left[C_F\frac{\alpha_s}{4\pi}\right]
\end{eqnarray}

Since one has 
$(\uparrow \downarrow)_{Born}^{pc}=(\downarrow \uparrow)_{Born}^{pc}
=2N_{c}q^{2}$ and 
$(\uparrow \downarrow)_{Born}^{pv}=-(\downarrow \uparrow)_{Born}^{pv}
=2N_{c}q^{2}$ one obtains the anomalous contributions in Eq.(\ref{hadfun}) 
using the factorizing relations (\ref{anom}). In particular one sees that
anomalous contribution to the single-spin polarization $P^{(\ell_{1})}$ 
results to
$100 \%$ from the universal helicity non-flip contribution whereas the 
anomalous contributions to the spin-spin correlation function
$P^{(\ell_{1}\ell_{2})}$ is $50\%$ helicity flip and $50\%$ helicity non-flip.
We conclude that the
anomalous non-flip contribution are unavoidable in the $O(\alpha_{s})$ 
description of $m_{q}\to0$ spin phenomena in $e^+e^-\to q \bar{q}\,(g)$.
The normal contributions in
Eq.(\ref{hadfun}) require an explicit calculation. The result
$H=4N_{c}q^{2}(1+\alpha_{s}/\pi)$ is, of course, well known since many
years.

In this talk we did not discuss  
physics aspects of the anomalous helicity flip contribution. We refer the 
interested reader to the papers
\cite{Smilga:1990uq,Falk:1993tf,Teryaev:1995eb,Teryaev:1996nf} which contain
a discussion of various aspects of the physics of the anomalous helicity flip 
contributions in QED and QCD.

\vspace*{7.2mm}\noindent
{\bf Acknowledgements:} The work of S.~G. is supported by the Estonian
target financed project No.~0180056s09, by the Estonian Science Foundation
under grant No.~8402 and by the Deutsche Forschungsgemeinschaft (DFG)
under grant 436 EST 17/1/06.

\end{document}